\documentclass{ws-p8-50x6-00}

\begin{document}
\newcommand{\dlq}{\lq\lq}
\newcommand{\ben}{\begin{eqnarray*}}
\newcommand{\een}{\end{eqnarray*}}
\newcommand{\stackeven}[2]{{{}_{\displaystyle{#1}}\atop\displaystyle{#2}}}
\newcommand{\lsim}{\stackeven{<}{\sim}}
\newcommand{\gsim}{\stackeven{>}{\sim}}
\newcommand{\un}[1]{\underline{#1}}
\renewcommand{\baselinestretch}{1.0}
\newcommand{\as}{\alpha_s}
\def\eq#1{{Eq.~(\ref{#1})}}
\def\fig#1{{Fig.~\ref{#1}}}

\title{High Energy QCD and the Large $N_c$ Limit}

\author{Yuri V. Kovchegov}

\address{Department of Physics, University of Washington, Box 351560 \\ 
Seattle, WA 98195 \\E-mail: yuri@phys.washington.edu}

%%%%%%%%%%%%%%%%%%%%%%%%%%%%%%%%%%%%%%%%%%%%%%%%%%%%%%%%%%%%%%
% You may repeat \author \address as often as necessary      %
%%%%%%%%%%%%%%%%%%%%%%%%%%%%%%%%%%%%%%%%%%%%%%%%%%%%%%%%%%%%%%

\maketitle

\begin{flushright}
\vspace*{-4.5cm}
NT@UW--02--003 \\
\vspace*{4.5cm}
\end{flushright}

\abstracts{We review recent progress in understanding QCD 
at high energies and the role played in it by the large $N_c$
limit. We discuss unitarization of total hadronic cross sections and
saturation of structure functions at high energies. }

\section{The BFKL Equation}

One of the most important problems in QCD is the problem of
understanding its high energy asymptotics. High energy QCD has direct
relevance to the scattering data produced at the accelerators
worldwide allowing for experimental verification of theoretical
attempts to understand QCD dynamics. In addition to that gluonic
fields in this extreme regime of high energies become very strong
making the underlying description of QCD processes non-perturbative,
while leaving the strong coupling constant $\as$ small. Therefore
understanding the gluon dynamics and correct degrees of freedom of QCD
at high energies is possible and may shed some light on the
non-perturbative dynamics of the theory.

Let us consider a scattering of two heavy quarkonia (quark--antiquark
states), or, equivalently, scattering of two virtual photons with
large virtualities $Q_1^2$ and $Q_2^2$, each of which splits into a
quark--antiquark pair. Large virtualities or large masses of the
quarkonia insure that the strong coupling constant is small, $\alpha_s
(Q_1^2 ) \sim \alpha_s (Q_2^2 ) \ll 1$, allowing us to use
perturbative expansion.

At the lowest order in $\alpha_s$ the simplest model of high energy
interaction between the two quarkonia is given by a two gluon exchange
in the forward amplitude. The corresponding total cross section of
onium--onium scattering is independent of energy at very high energies
and could be written as
\be
\sigma \, \sim \, s^0,
\ee
where $s$ is the center of mass energy of the system. This model of
high energy interactions is known as Low-Nussinov pomeron.\cite{ln} It
demonstrates that exchanging two vector particles (gluons) between the
onia gives a non-vanishing cross section at high energy, while using
scalars or fermions (with non-derivative couplings) would give a cross
section that decreases with energy.

However, at the high energies achievable by the modern days
accelerators the hadronic cross sections are not constant. They tend
to increase as powers of center of mass energy of the system
\be
\sigma \, \sim \, s^\Delta
\ee
where $\Delta$ is a number called pomeron intercept. The rise of
partonic structure functions with energy could be described by the
DGLAP evolution equation.\cite{dglap} Nevertheless the DGLAP equation
does not yield us with a power of energy growth of the cross
sections. The only way to obtain a cross section in perturbative QCD
that grows as a power of energy is given by the BFKL evolution
equation.\cite{bfkl}

\begin{figure}[t]
\begin{center}
\epsfxsize=5.5cm
\leavevmode
\hbox{ \epsffile{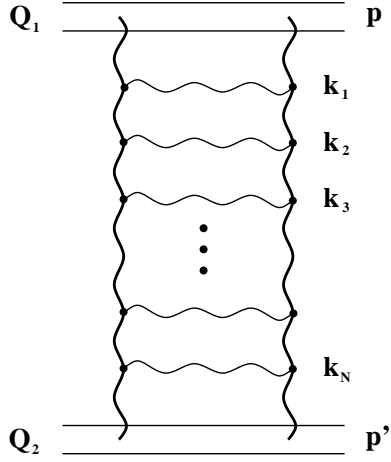}}
\end{center}
\caption{Ladder diagrams summed by the BFKL evolution equation. }
\label{ladder}
\end{figure}

The BFKL equation \cite{bfkl} resums ladder diagrams in the forward
onium--onium scattering amplitude of the type shown in
\fig{ladder}. The light cone momenta components of the $s$-channel
gluons are ordered (Regge kinematics) such that
\be
p_+ \, \gg \, k_{1+} \, \gg \, k_{2+} \, \ldots \, \gg \, k_{N+} \,
\gg \, p'_+
\ee
and
\be
p_- \, \ll \, k_{1-} \, \ll \, k_{2-} \, \ldots \, \ll \, k_{N-} \,
\ll \, p'_-
\ee
while there is no restrictions on their transverse momenta which
remain of the same order throughout the ladder
\be
Q_1 \, \sim \, k_{1\perp} \, \sim \, k_{2\perp} \, \sim \, \ldots \,
\sim k_{N\perp} \, \sim \, Q_2 .
\ee
Each $s$-channel gluon in the ladder brings in an extra power of the
coupling constant $\as \ll 1$. The integration over the longitudinal
phase space of each gluon gives a term proportional to $\ln s \gg
1$. Combining the two factors we see that each extra $s$-channel gluon
introduces a factor of $\as \ln s \, \sim \, 1$. Thus, by resumming
ladders of the type shown in \fig{ladder} the BFKL equation resums all
powers of the effective parameter $\as \ln s$.

By drawing a ladder in \fig{ladder} we are oversimplifying the
problem. The triple gluon vertices in the ladder are not the usual QCD
triple gluon vertices. These are the so-called effective Lipatov
vertices and have to be calculated by analyzing several gluon emission
diagrams\cite{bfkl}, though in certain gauges they do correspond to
the usual triple gluon vertices. The gluon propagators in the
t-channel are also not just the usual gluon propagators. They include
virtual (reggeization) corrections to the ladder which modify the
gluon propagators. The resulting reggeized gluons are denoted by thick
lines in \fig{ladder}, each line representing a summation of the whole
ladder of \fig{ladder} in the color octet channel.

The BFKL equation could be written for onium-onium scattering
amplitude $f (Q_1, Q_2, Y)$ as \cite{bfkl,mucar,jar}
\be\label{bint}
\frac{\partial f (\underline{l},\underline{l}', Y)}{\partial Y} 
\, = \, \int d^2 k \, K_{BFKL} (\underline{l},\underline{k}) \,
f (\underline{k},\underline{l}', Y)
\ee
where $Y = \ln (s/Q_1 Q_2)$ is the rapidity variable and $K_{BFKL}$ is
the kernel of the equation. Defining the Laplace-transformed amplitude
as
\be
f_\omega (\underline{l},\underline{l}') \, = \, \int_0^\infty \, dY \,
e^{- \omega Y} \, f (\underline{l},\underline{l}', Y)
\ee
\eq{bint} could be rewritten as \cite{bfkl,mucar,jar}
\be\label{bfkleq}
\omega f_{\omega} (\underline{l},\underline{l}') =  \, 
\delta^2 (\underline{l} - \underline{l}') + \frac{\alpha N_c}{\pi^2} 
\int \! \frac{d^2 k}{(\underline{k} - \underline{l})^2} \left[ f_{\omega}
(\underline{k},\underline{l}') - \frac{l^2 f_{\omega}
(\underline{l},\underline{l}')}{k^2 + (\underline{k} -
\underline{l})^2} \right].
\ee
The solution of \eq{bfkleq} in the saddle point approximation at large
$Y$ yields the scattering amplitude
\be\label{bfklsol}
f (Q_1, Q_2, Y) \, = \, \frac{1}{2\pi Q_1 Q_2 \sqrt{4\pi D Y}} \,
e^{(\alpha_P - 1 ) Y} \, \exp \left[-{\ln^2 Q_1/Q_2 \over 4 \, D \,
Y}\right]
\ee
with $D = {7 \as \over 2\pi} N_c \zeta(3)$ and $\alpha_P - 1 = {4\as
N_c \over \pi} \ln 2$. At very high energy the solution given by
\eq{bfklsol} grows as a power of energy since
\be\label{poe}
e^{(\alpha_P - 1 ) Y} \, \sim \, s^{\alpha_P - 1}.
\ee
Therefore the BFKL equation seems to be a natural candidate for the
phenomenological pomeron as observed in high energy scattering
experiments. It is beyond the scope of this talk to address the issue
of how well the BFKL equation describes the data. We just note that
the numerical value of the obtained pomeron intercept $\alpha_P - 1 =
{4\as N_c \over \pi} \ln 2 \, \approx \, 2.6 \, \as \, \approx \, 0.8$
is higher than the experimental value of $0.2 - 0.3$ observed in deep
inelastic scattering experiments. The inclusion of next-to-leading
order correction to BFKL kernel makes the intercept negative for all
reasonable values of $\as$ \cite{nlo,nlo1,nlo2}, though this problem
is likely to be due to collinear singularities and should be cured at
higher orders in $\as$ \cite{ccs} resulting in a slightly lower but
still positive value of the intercept which is closer to experimental
data. The BFKL equation as described here is applicable only to the
scattering of perturbative (small) objects like heavy quarkonia and
can not be applied to the total cross section in, say, proton--proton
collisions, where (in the single ladder approximation) there is no
hard scale which would justify the small coupling approach.

However, the BFKL equation poses some important questions even in the
case of heavy onium-onium scattering and at the leading order in $\as$
in the kernel.

{\bf (i)} The power of energy growth of the total cross section
(\eq{poe}) violates Froissart unitarity bound, which states that the
growth of the total cross sections with energy at asymptotically high
energies is bounded by \cite{froi}
\be
\sigma \, \le \, \frac{const}{m_\pi^2} \, \ln^2 s
\ee
with $m_\pi$ the pion's mass. (For a good pedagogical derivation of
Froissart bound I refer the readers to \cite{genya}.) That means some
new effects should modify the BFKL equation at very large $s$ to make
the resulting amplitude unitary.

{\bf (ii)} The solution of \eq{bfklsol} includes a diffusion term. To
see this let us consider a half of the ladder, from one of the onia to
some intermediate gluon in the middle of the ladder carrying momentum
$k_\perp$ and rapidity $Y/2$. Applying \eq{bfklsol} to that
half-ladder we see that it includes a term
\be\label{diff}
\exp \left[-{\ln^2 k_\perp/Q \over 2 \, D \, Y}\right].
\ee
This term is responsible for diffusion of the transverse momenta from
the initial perturbative scale $Q$ both into infrared and
ultraviolet. The width of the diffusion grows with rapidity $Y$. Thus
no matter how large the starting scale $Q$ is at certain very high
energy the momentum of some gluons in the middle of the ladder would
become of the order of $\Lambda_{QCD}$ invalidating further
application of BFKL evolution.\cite{bartels} Again this hints that the
BFKL equation should be modified at higher energies to avoid this
problem.

\section{Multiple Reggeon Exchanges}

\begin{figure}[b]
\begin{center}
\epsfxsize=7cm
\leavevmode
\hbox{ \epsffile{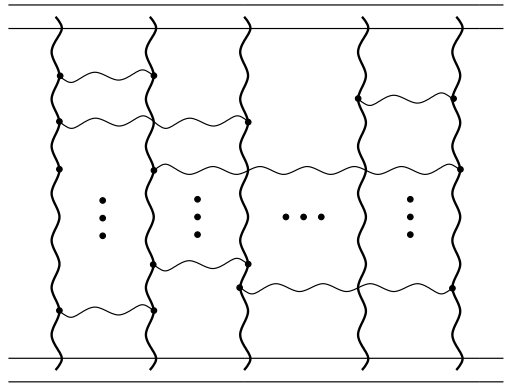}}
\end{center}
\caption{Multiple reggeon exchange diagram.}
\label{multreg}
\end{figure}

The BFKL ladder in \fig{ladder} can be viewed as two reggeized gluons
(reggeons) propagating in the $t$-channel. It has been conjectured
that unitarity of the total cross section could be restored if one
resums all diagrams with multiple reggeon exchanges
\cite{mult}. Instead of summing the ladder in \fig{ladder} one has to
analyze all the diagrams of the type shown in \fig{multreg}. There one
can exchange any number of reggeons in the $t$-channel. Each pair of
reggeons may interact via $s$-channel gluon exchanges leading to
BFKL-type contributions. To complete the picture one should also
include the diagrams where the number of reggeons varies in the
$t$-channel, that is the diagrams containing vertices of $n$ reggeons
going into $m$ reggeons for any $n$ and $m$. This is beyond the scope
of this talk. We are going to concentrate on the diagrams where the
number of reggeons is constant.

The $n$-reggeon exchange diagrams were first resummed by Matinyan and
Sedarkyan \cite{matsed} for the scalar $\phi^3$ theory. For
$n$-reggeon exchange amplitude one can write down an equation similar
to \eq{bint}. However there are many more contributions to the kernel
than in the $2$-reggeon case considered above. Since at each step in
rapidity any two out of $n$ reggeons can interact one expects $n (n-1)
/2$ contributions to the kernel of the integral equation. Each
contribution comes in with the same integral kernel acting on a
different pair of reggeons. Assuming for simplicity that all these
kernels are approximately the same we can write the following equation
for $n$-reggeon amplitude ${\cal F}_n$ at large $n$
\be\label{nf1}
\frac{\partial {\cal F}_n}{\partial Y} \, \approx \, n^2 \, 
K \otimes {\cal F}_n.
\ee
The solution of \eq{nf1} scales as
\be
{\cal F}_n (Y) \, \sim \, e^{(- 2 n + \mbox{const} \, n^2) Y} \,
\sim \, s^{- 2 n + \mbox{const} \, n^2},
\ee
where the term $- 2 n$ in the power results from the fact that we are
dealing with $n$ scalar particles in $\phi^3$ theory. The total cross
section would be given by the sum of $n$-reggeon amplitudes
\be\label{sst}
\sigma \, \sim \, \sum_n \, {\cal F}_n \, \sim \, \sum_n \, 
c_n \, s^{- 2 n + \mbox{const} \, n^2},
\ee
which is not likely to converge! ($c_n$'s are some coefficients which
may depend on $s$ at most logarithmically.) This result indicates that
multiple reggeon exchanges in $\phi^3$ theory not only fail to
unitarize the total cross section, but rather make the series highly
divergent. It is believed that the problem is associated to the fact
that $\phi^3$ theory has no stable vacuum and is typical only to
$\phi^3$ theory. Nevertheless one might and should get worried about
similar phenomenon happening in QCD.

\begin{figure}[t]
\begin{center}
\epsfxsize=6cm
\leavevmode
\hbox{ \epsffile{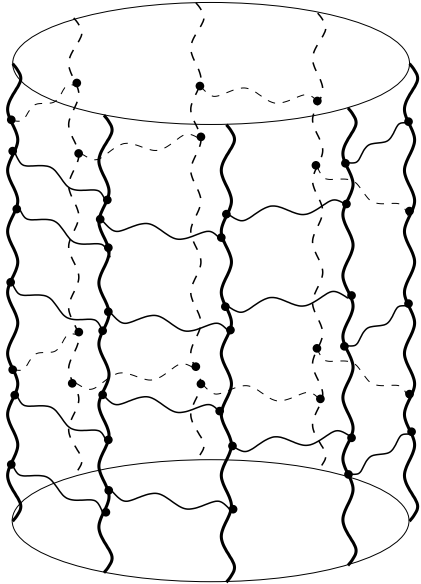}}
\end{center}
\caption{An example of cylinder topology acquired by multiple reggeon 
exchange diagrams in the large $N_c$ limit. The ellipses are drawn to
guide the eye. }
\label{cyl}
\end{figure}

To demonstrate that the catastrophe which happened in $\phi^3$ theory
is actually avoided in QCD one has to invoke the 't Hooft large $N_c$
limit \cite{tht} which is the central topic of this conference. In the
large $N_c$ limit the diagrams in \fig{multreg} drastically simplify:
only the planar graphs survive. Thus the topology of the interaction
takes form of a cylinder with the reggeons forming the walls of the
cylinder and only the interactions of the nearest neighbor reggeons
are allowed, as depicted in \fig{cyl}. There the $i$th reggeon may
interact only with the $i-1$st and $i+1$st reggeons, and the $1$st
reggeon interacts with the $n$th and the $2$nd reggeons. The two
reggeon case (the pomeron of \fig{ladder}) corresponds to the
simplest cylinder topology case with reggeons and $s$-channel gluons
``living'' on the walls of a cylinder. If we add one more reggeon we
would obtain the three reggeon case (odderon) which also has a
cylindrical topology as shown in \fig{cyl}.

With only the nearest neighbor interactions between the reggeons
allowed the kernel of the integral equation resumming the diagrams
shown in \fig{cyl} receives only $n$ contributions, which is very
different from $\phi^3$ theory. Making the same assumption as for
$\phi^3$ theory we see that \eq{nf1} for large-$N_c$ QCD becomes
\be\label{fn1}
\frac{\partial {\cal F}_n}{\partial Y} \, \approx \, n \, 
K \otimes {\cal F}_n
\ee
leading to 
\be\label{fv}
{\cal F}_n (Y) \, \sim \, e^{\mbox{const} \, n \, Y} \,
\sim \, s^{\mbox{const} \, n}.
\ee
Now the total cross section similar to \eq{sst} is
\be\label{svt}
\sigma \, \sim \, \sum_n \, {\cal F}_n \, \sim \, \sum_n \, 
d_n \, s^{\mbox{const} \, n},
\ee
with $d_n$'s some coefficients with weak dependence on $s$. The series
in \eq{svt} could be convergent and therefore large-$N_c$ QCD avoids
the catastrophe of $\phi^3$ theory. One expects that the problem would
not reappear for the real life case of three colors. We have to advise
the reader to understand \eq{fv} only as an upper bound on the $n$
reggeon amplitudes but not as an exact answer.

A careful QCD analysis of the reggeon cylinder graphs of the type
shown in \fig{cyl} demonstrates that a reggeon system is equivalent to
a one dimensional XXX Heisenberg spin chain for spin $S = 0$ and is
thus an exactly solvable model \cite{heis}. That facilitates the
analysis of the $n$-reggeon diagrams allowing to quantitatively
understand their high energy behavior.\cite{korch} 

An important question for the total cross section is whether the
leading-$N_c$ $2 n$-reggeon diagram gives an amplitude which grows
with energy faster than the subleading in $N_c$ $n$-pomeron
diagram. Naturally the $n$-pomeron diagram should scale as $s^{n
(\alpha_P - 1)}$. In the recent paper by Korchemsky et al\cite{korch}
it has been shown that the $n$ reggeon amplitude seem to scale as
${\cal F}_n \, \sim \, s^{\mbox{const}/n}$. Thus at very high energies
multiple reggeon exchange diagrams grow much slower than multiple
pomeron exchange diagrams and the former could be neglected compared
to the latter. We refer the interested readers to the contribution of
G. Korchemsky in these proceedings \cite{korch}.

\section{Multiple Pomeron Exchanges}

\subsection{General Physical Picture}

As could be seen from the discussion in the previous section an
exchange of a single BFKL pomeron brings in a parametrical factor of
$\as^2 \, e^{(\alpha_P -1) Y}$, where $e^{(\alpha_P -1) Y}$ comes from
the amplitude in \eq{bfklsol} and the factor of $\as^2$ comes from
connecting the reggeons in the ladder to the onia. The BFKL equation
resums powers of $(\alpha_P - 1) Y \, \sim \, \as Y$ and therefore
gives a large contribution when
\be
Y_{LO} \, \sim \, \frac{1}{\as}.
\ee
What happens as we increase the energy/rapidity? It is possible that
multiple pomeron exchanges would start playing a bigger role in the
scattering. To estimate when this happens we have to equate single
($\as^2 \, e^{(\alpha_P -1) Y}$) and double ($\as^4 \, e^{2 (\alpha_P
-1) Y}$) pomeron contributions, effectively requiring that $\as^2 \,
e^{(\alpha_P -1) Y_U} \, \sim \, 1$. As was argued by Mueller in
\cite{pl:b396:251} multiple pomeron exchanges become important at the
values of rapidity of the order of
\begin{equation}\label{eq:yvk:yu}
  Y_U \sim \frac{1}{\alpha_P - 1} \, \ln \frac{1}{\as^2}.
\end{equation}

After completion of the calculation of the next--to--leading order
corrections to the kernel of the BFKL equation (NLO BFKL) by Fadin and
Lipatov \cite{nlo}, and, independently, by Camici and Ciafaloni
\cite{nlo1}, it was shown in \cite{nlo2} that due to the running
coupling effects these corrections become important at the rapidities
of the order of
\begin{equation}\label{ee:yvk:rc}
Y_{NLO} \sim \frac{1}{\as^{5/3}}.
\end{equation}
One can see that $Y_{LO} \, \ll \, Y_U \, \ll \, Y_{NLO}$ for
parametrically small $\as$.  That implies that the center of mass
energy at which the multiple pomeron exchanges become important is
much smaller, and, therefore, is easier to achieve, than the energy at
which NLO corrections start playing an important role. That is
multiple pomeron exchanges are probably more relevant than NLO BFKL to
the description of current experiments.\cite{epj:c7:609}  Also
multiple pomeron exchanges are much more likely to unitarize the total
hadronic cross section. Here we are going to present a solution to the
problem of resummation of multiple pomeron exchanges and show how the
BFKL pomeron unitarizes following\cite{me}.

\begin{figure}
\begin{center}
\epsfxsize=7cm
\leavevmode
\hbox{ \epsffile{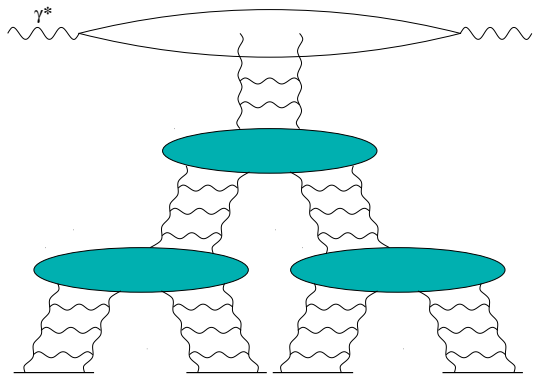}}
\end{center}
\caption{An example of the pomeron ``fan'' diagram. Each ladder interacts 
with the proton or nucleus at rest independently.}
\label{fig:yvk:fan}
\end{figure}

We will consider deep inelastic scattering (DIS) of a virtual photon
on a hadron or nucleus and will resum all multiple pomeron exchanges
contributing to the total DIS cross section in the leading
longitudinal logarithmic approximation in the large $N_c$ limit. The
first step in that direction in perturbative QCD was a conjecture by
Gribov, Levin and Ryskin (GLR)\cite{glr} of an equation describing the
fusion of two pomeron ladders into one, which was proven in the double
logarithmic limit by Mueller and Qiu.\cite{mq} The resulting equation
resums all pomeron ``fan'' diagrams (see \fig{fig:yvk:fan}) in the
double logarithmic approximation. An extensive work on resumming the
multiple pomeron exchanges in the gluon distribution function in the
leading $\ln (1/x)$ approximation (i.e. without taking the double
logarithmic limit) both in the framework of effective field theories
and employing alternative approaches has been pursued in
\cite{jklw,ilm,np:b463:99,np:b493:305}.

\begin{figure}
\begin{center}
\epsfxsize=10cm
\leavevmode
\hbox{ \epsffile{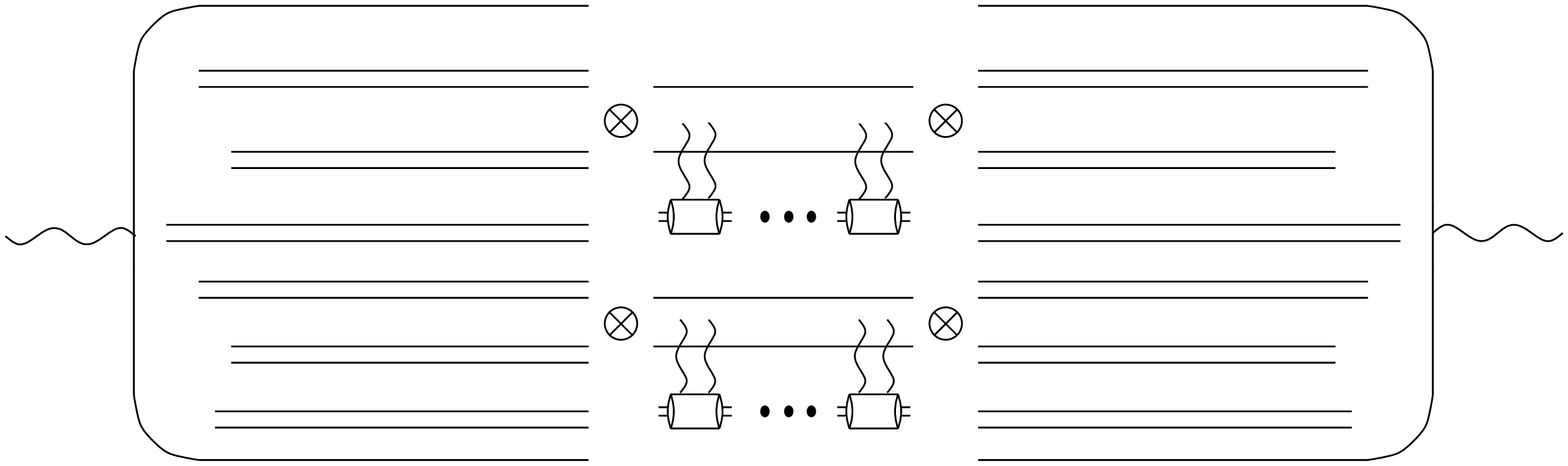}}
\end{center}
\caption{Dipole evolution in the deep inelastic scattering process as
  pictured in\protect\cite{me}. The incoming virtual photon develops a
  system of color dipoles, each of which rescatters on the target
  hadron or nucleus (only two are shown). }
\label{fig:yvk:dis}
\end{figure}

Let us consider a deep inelastic scattering (DIS) of a virtual photon
on a hadron or nucleus. An incoming photon splits into a
quark--antiquark pair and then the $q{\overline q}$ pair rescatters on
the target hadron or nucleus. In the rest frame of the hadron all QCD
evolution should be included in the wave function of the incoming
photon. That way the incoming photon develops a cascade of gluons,
which then scatter on the hadron at rest. We want to calculate this
gluon cascade in the leading longitudinal logarithmic ($\ln 1/x$)
approximation and in the large $N_c$ limit.  This is exactly the type
of cascade described by Mueller's dipole model\cite{dip}. In the large
$N_c$ limit the incoming gluon develops a system of color dipoles and
each of them independently rescatters on the hadron or nucleus. The
forward amplitude of the process is shown in
\fig{fig:yvk:dis}. The double lines in \fig{fig:yvk:dis}
correspond to gluons in large $N_c$ approximation being represented as
consisting of a quark and an antiquark of different colors. The color
dipoles are formed by a quark from one gluon and an antiquark from
another gluon. In \fig{fig:yvk:dis} each dipole, that was developed
through the QCD evolution, later interacts with the hadron or nucleus
by a series of Glauber--type multiple rescatterings on the sources of
color charge.  The assumption about the type of interaction of the
dipoles with the hadron is not important for the evolution. It could
also be just two gluon exchanges. The important assumption is that
each dipole interacts with the target independently of the other
dipoles, which is done in the spirit of the large $N_c$ limit and is
valid for a model of a large dilute target nucleus.

For the case of a dilute target independent dipole interactions with
the target proton or nucleus are enhanced by some factors of the
number of sources of color charge in the proton or nucleus compared to
the case when several dipoles interact with the same parton in the
proton or the same nucleon in the nucleus.\cite{mv,yuri} This allowed
us to assume that the dipoles interact with the proton or nucleus
independently. In case of a nucleus one actually has the atomic number
$A$ as a parameter leading powers in which justify this
approximation.\cite{mv,yuri} The summation of these contributions
corresponds to summation of the pomeron ``fan'' diagrams of
Fig. \ref{fig:yvk:fan}. In general there is another class of diagrams,
which we will call pomeron loop diagrams.  In a pomeron loop diagram a
pomeron first splits into two pomerons, just like in the fan diagram,
but then the two pomerons merge back into one pomeron. The graphs
containing pomerons not only splitting, but also merging together will
be referred to as pomeron loop diagrams. In a dilute proton or nucleus
these pomeron loop diagrams could be considered small, since they
would be suppressed by powers of color charge density ($A$).  At very
high energies this approximation becomes not very well justified even
in the dilute target case.  Contribution of each additional pomeron in
a fan diagram on a hadron is parametrically of the order of $\as^2
A^{1/3} e^{(\alpha_P - 1)Y}$ which is large at the rapidities of the
order of $Y_U$ given by Eq. (\ref{eq:yvk:yu}). The contribution of an
extra pomeron in a pomeron loop diagram is of the order of $\as^2
e^{(\alpha_P - 1)Y}$, which is not enhanced by powers of $A$ and is,
therefore, suppressed in the dilute nucleus case. However, one can
easily see that at extremely high energies, when $\as^2
e^{(\alpha_P - 1)Y}$ becomes greater than one, pomeron loop diagrams
become important, still being smaller than the fan
diagrams. Resummation of pomeron loop diagrams in a dipole wave
function seems to be a very hard technical problem, possibly involving
NLO BFKL or, even, next-to-next-to-leading order BFKL kernel
calculations.\cite{np:b507:367} Nevertheless, it is the author's
belief that while being crucial for onium--onium scattering the
pomeron loop diagrams are not going to significantly change the high
energy behavior of the DIS structure functions.

\subsection{Nonlinear Evolution Equation}

We start by considering a deep inelastic scattering process. The
incoming virtual photon with a large $q_+$ component of the momentum
splits into a quark--antiquark pair which then interacts with the
target at rest.  We model the interaction by no more than two gluon
exchanges between each source of color charge and the quark--antiquark
pair. This is done in the spirit of the quasi--classical approximation
used previously in
\cite{np:b335:115,np:b529:451}. The interactions are taken in the eikonal
approximation. Then, as could be shown in general, i.e., including the
leading logarithmic QCD evolution, the total cross--section, and,
therefore, the $F_2$ structure function of the nucleus can be
rewritten as a product of the square of the virtual photon's wave
function and the propagator of the quark--antiquark pair through the
proton. The expression reads
\begin{equation}\label{eq:yvk:defN}
  F_2 (x, Q^2) = \frac{Q^2}{4 \pi^2 \alpha_{EM}} \int \frac{d^2 {\bf
      x}_{01} d z }{2 \pi} \, [\Phi_T ({\bf x}_{01},z) + \Phi_L ({\bf
    x}_{01},z) ] \ d^2 b_0 \ N({\bf x}_{01},{\bf b}_0 , Y) ,
\end{equation}
where the incoming photon with virtuality $Q$ splits into a
quark--antiquark pair with the transverse coordinates of the quark and
antiquark being ${\bf \tilde{x}}_0$ and ${\bf \tilde{x}}_1$
correspondingly, such that ${\bf x}_{10} = {\bf \tilde{x}}_1 - {\bf
\tilde{x}}_0$. The coordinate of the center of the pair is given by
${\bf b}_0 = \frac{1}{2} ( {\bf\tilde{x} }_1 + {\bf \tilde{x}}_0 )$.
The square of the light cone wave function of $q \overline{q}$
fluctuations of a virtual photon is denoted by $\Phi_T ({\bf
x}_{01},z)$ and $\Phi_L ({\bf x}_{01},z)$ for transverse and
longitudinal photons correspondingly, with $z$ being the fraction of
the photon's longitudinal momentum carried by the quark. $\Phi_T ({\bf
x}_{01},z)$ and $\Phi_L ({\bf x}_{01},z)$ are known functions and
could be found in the literature.\cite{zp:c49:607}

The quantity $N({\bf x}_{01},{\bf b}_0, Y)$ has the meaning of the
forward scattering amplitude of the quark--antiquark pair on a
nucleus. At the lowest (classical) order not including the QCD
evolution in rapidity it is given by
\begin{equation}\label{eq:yvk:gamma2}
  N({\bf x}_{01},{\bf b}_0, 0) = - \gamma ({\bf x}_{01},{\bf b}_0)
  \approx \left\{ 1 - \exp \left[ - \frac{\as \pi^2}{2 N_c S_\perp}
  {\bf x}_{01}^2 A \, xG ( x ,1/{\bf x}_{01}^2 ) \right] \right\}.
\end{equation}
$\gamma ({\bf x}_{01},{\bf b}_0)$ is the propagator of the $q
\overline{q}$ pair through the hadron or nucleus. The propagator 
could be easily calculated, similarly to
\cite{np:b335:115,np:b529:451}, giving the Glauber multiple
rescattering formula (\ref{eq:yvk:gamma2}). Throughout the talk we
assume for simplicity that the target nucleus is a cylinder, which
appears as a circle of radius $R$ in the transverse direction and has
a constant length $2 R$ along the longitudinal $z$ direction.
Therefore its transverse cross--sectional area is $S_\perp = \pi R^2$.
In formula (\ref{eq:yvk:gamma2}) $\as$ is the strong coupling constant
and $xG ( x ,1/{\bf x}_{01}^2 )$ is the gluon distribution of a
nucleon in the nucleus, taken at the lowest order in $\as$, similarly
to \cite{np:b529:451}.

Eq. (\ref{eq:yvk:gamma2}) resums all Glauber type multiple
rescatterings of a $q \overline{q}$ pair on a target nucleus. As was
mentioned before, since each interaction of the pair with a source of
color charge in the nucleus is restricted to the two gluon exchange,
the formula (\ref{eq:yvk:gamma2}) effectively sums up all the powers
of the parameter $\as^2 A^{1/3}$. (Since $xG \sim C_F$ there is no
$N_c$ present in the expansion in the large-$N_c$ limit.) Therefore
this interaction of course is subleading in $N_c$ as is true for any
hadronic interactions.\cite{tht}

Since the target is at rest in order to include the QCD evolution of
$F_2$ structure function, we have to develop the soft gluon wave
function of the incoming virtual photon. In the leading longitudinal
logarithmic approximation ($\ln 1/x$) the evolution of the wave
function is realized through successive emissions of small--$x$
gluons.  The $q \overline{q}$ pair develops a cascade of gluons, which
then scatter on the hadron. In order to describe the soft gluon
cascade we will take the limit of a large number of colors, $N_c
\rightarrow \infty$. Then, this leading logarithmic soft gluon
wavefunction will become equivalent to the dipole wave function,
introduced by Mueller in \cite{dip}. The physical picture becomes
straightforward. The $q \overline{q}$ pair develops a system of
dipoles (dipole wave function), and each of the dipoles independently
scatters on the target hadron or nucleus (see
\fig{fig:yvk:dis}). Since the target is assumed to have many sources
of color charge in it we may approximate the interaction of a dipole
(quark--antiquark pair) with it by $\gamma ({\bf x},{\bf b})$ given by
Eq.  (\ref{eq:yvk:gamma2}), with ${\bf x}$ and ${\bf b}$ being the
dipole's transverse separation and impact parameter. That means that
each of the dipoles interacts with several sources of color charge
(Glauber rescattering) in the nucleus independent of other dipoles.

\begin{figure}
\begin{center}
\epsfxsize=11cm
\leavevmode
\hbox{ \epsffile{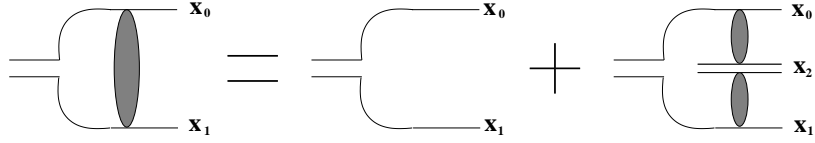}}
\end{center}
\caption{A diagrammatic representation of the dipole evolution equation.}
\label{fig:yvk:dipeqn}
\end{figure}

To construct the dipole wave function we will heavily rely on the
techniques developed in \cite{dip}.  We have to resum a gluonic
cascade in the large-$N_c$ limit. Denoting the cascade by a blob we
can write down an equation which is illustrated in
\fig{fig:yvk:dipeqn}.  As usual for the large $N_c$ limit the double 
line denotes a gluon.  The original $q\bar q$ pair may have no
evolution in it, i.e., the cascade could be empty, as illustrated by
the first term on the right hand side of \fig{fig:yvk:dipeqn}. In one
step of the evolution in rapidity a gluon is emitted in a dipole,
splitting the parent dipole into two dipoles. Each of the produced
dipoles is nonlocal: it consists of a quark (antiquark) from the
original quark--antiquark pair and an antiquark (quark) component of
the emitted gluon. The consecutive gluon evolution can continue in
either of the two dipoles. This is depicted in the second term on the
right hand side of \fig{fig:yvk:dipeqn}. This procedure gives us a
self-consistent equation resumming the gluon cascade.\cite{dip,me}

Substituting $1 - N({\bf x}_{01},{\bf b}_0, Y)$ instead of the blob
symbolizing the gluon cascade and including virtual corrections one
obtains the following equation\cite{me}
\begin{eqnarray*}
  N({\bf x}_{01},{\bf b}_0, Y) = - \gamma ({\bf x}_{01},{\bf b}_0) \,
  \exp \left[ - \frac{4 \as C_F}{\pi} \ln \left(
      \frac{x_{01}}{\rho} \right) Y \right]  
\end{eqnarray*}  
\begin{eqnarray*}
+ \frac{\as C_F}{\pi^2}
  \int_0^Y d y \, \exp \left[ - \frac{4 \as C_F}{\pi} \ln \left(
      \frac{x_{01}}{\rho} \right) (Y - y) \right]
\end{eqnarray*}  
\begin{eqnarray*}
  \times \int_\rho d^2 \tilde{x}_2 \frac{x_{01}^2}{x_{02}^2 x_{12}^2}
  \, [ 2 \, N({\bf x}_{02},{\bf b}_0 + \frac{1}{2} {\bf x}_{12}, y)
\end{eqnarray*}  
\begin{eqnarray}\label{eqN}
- N({\bf x}_{02},{\bf b}_0 + \frac{1}{2} {\bf x}_{12}, y) \, N({\bf
    x}_{12},{\bf b}_0 - \frac{1}{2} {\bf x}_{20}, y) ] .
\end{eqnarray}
Equation (\ref{eqN}), together with Eq. (\ref{eq:yvk:defN}), provide
us with the leading logarithmic evolution of the $F_2$ structure
function of a hadron including all multiple pomeron exchanges in the
large--$N_c$ limit. Note that while the evolution is leading in $N_c$
the interactions ($\gamma$) are not, so that \eq{eqN} could be viewed
as resumming powers of $\as^2 A^{1/3} \, e^{\as N_c Y}$.

Similar evolution equation was obtained by Balitsky in
\cite{np:b463:99} using an effective lagrangian approach to high 
energy QCD. The evolution equations obtained in \cite{np:b463:99}
reduce to a closed single equation only in the large-$N_c$
limit. Similar results have been obtained in \cite{jklw,ilm}.

\subsection{Solution of the Nonlinear Evolution Equation}

To understand the qualitative features of the integral equation
(\ref{eqN}) it is convenient to rewrite it in the differential
form. Assuming that the typical size of the dipole wave function
$x_\perp \ll R$, where $R$ is the proton's radius, we can rewrite
\eq{eqN} as
\begin{eqnarray*}
  \frac{\partial N({\bf x}_{01}, Y)}{\partial Y} = \frac{2 \as
    C_F}{\pi^2} \, \int_\rho d^2 x_2 \left[ \frac{x^2_{01}}{x^2_{02}
      x^2_{12}} - 2 \pi \delta^2 ({\bf x}_{01} -{\bf x}_{02}) \ln
    \left( \frac{x_{01}}{\rho} \right) \right] N({\bf x}_{02}, Y)
\end{eqnarray*}
\begin{eqnarray}\label{eq:yvk:1diff}
  - \frac{\as C_F}{\pi^2} \, \int d^2 x_2 \,
  \frac{x^2_{01}}{x^2_{02} x^2_{12}} \, N({\bf x}_{02}, Y) \, N({\bf
    x}_{12}, Y).
\end{eqnarray}
Performing a Fourier transform 
\begin{eqnarray}\label{eq:yvk:trfm}
  N(x_\perp, Y) = x_\perp^2 \int \frac{d^2 k}{2 \pi} \, e^{i {\bf k}
    \cdot {\bf x}} \, {\tilde N} (k, Y) = x_\perp^2 \int_0^\infty dk
  \, k \, J_0 (k x_\perp) \, {\tilde N} (k, Y).
\end{eqnarray}
we obtain
\begin{eqnarray}\label{ksp}
  \frac{\partial {\tilde N} (k, Y)}{ \partial Y} = \frac{2 \as
    N_c}{\pi} \, \chi \left( - \frac{\partial}{\partial \ln k} \right)
  \, {\tilde N} (k, Y) - \frac{\as N_c}{\pi} \, {\tilde N}^2 (k,
  Y),
\end{eqnarray}
where
\begin{equation}\label{eq:yvk:chi}
  \chi (\lambda) = \psi (1) - \frac{1}{2} \psi \left( 1 -
    \frac{\lambda}{2} \right) - \frac{1}{2} \psi \left(
    \frac{\lambda}{2} \right)
\end{equation}
is the eigenvalue of the BFKL kernel \cite{bfkl} with $\psi (\lambda)
= \Gamma ' (\lambda) / \Gamma (\lambda)$. In Eq.  (\ref{ksp}) the
function $\chi (\lambda)$ is taken as a differential operator with
$\lambda = - \partial / \partial \ln k$ acting on ${\tilde N} (k,
Y)$. We also put $N_c / 2$ instead of $C_F$ everywhere in the spirit
of the large $N_c$ approximation. The details of obtaining
Eq. (\ref{ksp}) from Eq.  (\ref{eq:yvk:1diff}) could be found in
\cite{me}.

Now the physical meaning of the evolution given by \eq{eqN} becomes
manifest. The linear part of \eq{ksp} is just the familiar BFKL
equation, giving a cross section growing as a power of energy. When
energy is not too large and $N$ is small the quadratic term could be
neglected and the linear evolution is all that remains. However as the
energy increases the evolution driven by BFKL would make $N$ large,
making the quadratic term in \eq{ksp} important. The quadratic term
would introduce damping effects, similar to GLR-MQ
equation\cite{glr,mq}, slowing down the growth of the cross section
and possibly unitarizing it.

\begin{figure}
\begin{center}
\epsfxsize=10cm
\leavevmode
\hbox{ \epsffile{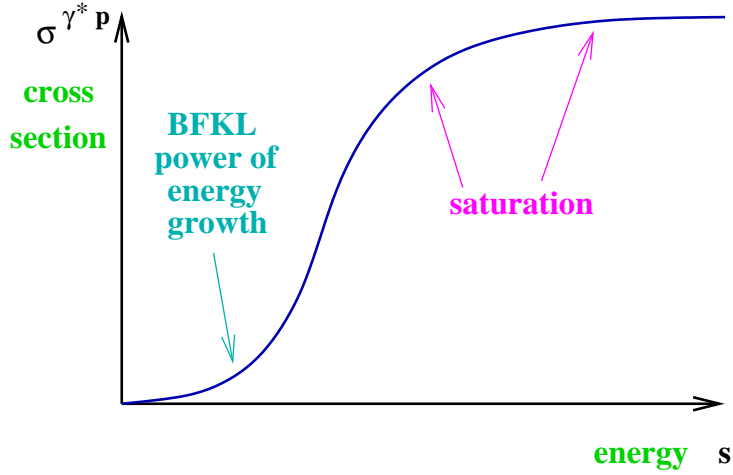}}
\end{center}
\caption{Qualitative behavior of the total DIS cross section given 
by the solution of \protect\eq{eqN}.}
\label{satur}
\end{figure}

\eq{eqN} has been solved both analytically by approximate methods 
\cite{me,lt} and numerically\cite{lt,braun,lub,gbm}. The result 
for DIS cross sections is qualitatively illustrated in
\fig{satur}. The cross section starts growing as a power of energy 
and then saturates to a much slower (logarithmic) growth.\cite{me,lt}
Thus \eq{eqN} resolves the problem of unitarization of BFKL pomeron
{\bf (i)} posed in Sect. 1.

Does \eq{eqN} answer the second question {\bf (ii)} posed in Sect. 1? 
Let us first note that \eq{eqN} produces a large momentum scale which
grows with energy as predicted within the framework of
McLerran-Venugopalan model\cite{mv,jklw,ilm}. To see this let us write
down the solution to \eq{ksp} in the limit of moderate energies (where
the equation is still mostly linear):
\begin{equation}\label{eq:yvk:nn1}
{\tilde N}_1 (k, Y) = C \, \frac{\Lambda}{k} \,
\frac{\exp [ (\alpha_P - 1)Y] }{\sqrt{14 \as N_c \zeta (3) Y}} 
\exp \left( - \frac{\pi}{14 \as N_c \zeta (3) Y} \ln^2 
\frac{k}{\Lambda} \right),
\end{equation}
where $C$ is some overall normalization constant and $\Lambda$ is the
momentum scale inherent to the initial distribution given by
\eq{eq:yvk:gamma2} (in fact it is the classical saturation scale 
given by McLerran-Venugopalan model\cite{mv,yuri,np:b529:451}). At
very high energy the amplitude $N$ saturates to a constant: $N
\rightarrow 1$ as $Y \rightarrow \infty$. In momentum space this 
corresponds to\cite{me} 
\begin{equation}\label{eq:yvk:log}
\tilde{N}_{sat} (k, Y) = \ln \frac{Q_s}{k}.
\end{equation}
Logarithm is a slowly varying function compared to exponent and to
estimate at what momentum scale the transition to saturation takes
place we can just equate the amplitude in \eq{eq:yvk:nn1} to
one. Omitting all slowly varying with $Y$ prefactors we obtain an
estimate of the the saturation scale
\begin{equation}\label{eq:yvk:satsc}
Q_s (Y) \, \approx \, \Lambda \, \frac{\exp [
(\alpha_P - 1)Y] }{\sqrt{14 \as N_c \zeta (3) Y}}. 
\end{equation}
One can see from \eq{eq:yvk:satsc} that as energy increases the
saturation scale increases too. 

Numerical simulation\cite{braun,lub,gbm} demonstrated a very
interesting behavior of the solution of \eq{eqN}, which is drastically
different from the solution of BFKL. Let us remind the readers that in
Sect. 1 based on \eq{diff} we observed that with increasing energy the
momentum of gluons given by the solution to BFKL equation diffuses
around some initial external perturbative scale $Q$ with the diffusion
width increasing with $Y$. That was potentially dangerous since the
diffusion cone would eventually get into the non-perturbative
region. The solution of \eq{eqN} produces a very different
distribution of transverse momenta of the gluons: the width of the
diffusion cone seems to be independent of energy while the center of
the distribution is given by the saturation scale $Q_s (Y)$ which
increases with $Y$\cite{braun,lub,gbm} (see \eq{eq:yvk:satsc}). Thus
the distribution never gets into the dangerous infrared
region. Therefore \eq{eqN} appears to solve the problem {\bf (ii)}
posed in Sect. 1 of the sensitivity of the cross sections given by
BFKL evolution to the non-perturbative infrared region.\cite{bartels}

\section{``Phase Diagram'' of High Energy QCD}

\begin{figure}[b]
\begin{center}
  \epsfxsize=9cm
\leavevmode
\hbox{ \epsffile{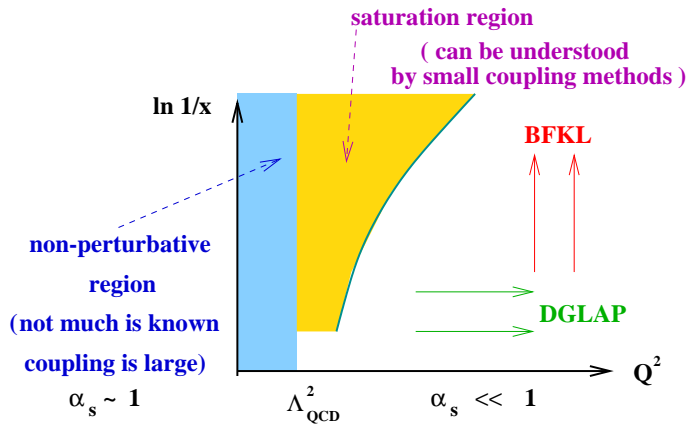}}
\end{center}
\caption{``Phase diagram'' of high energy QCD.}
\label{psat}
\end{figure}

To summarize the results of the previous Section we present the
``phase diagram'' of high energy QCD depicted in \fig{psat}.

The majority of the HERA data shows that the proton's structure
functions at large photon's virtuality $Q^2$ and not very small
Bjorken $x \, \approx \, Q^2/s$ can be described by the DGLAP
evolution equation \cite{dglap}, which is a linear equation. It is
convenient to present different properties of hadronic structure
functions in terms of a ``phase diagram'' in $Q^2$ and $\ln 1/x$ plane
(see Fig. \ref{psat}). The DGLAP physics corresponds to the lower
right section of the plane, where $Q^2$ is large and $x$ is not too
small. As one goes towards smaller $x$ in the same region of $Q^2$ the
hadronic structure functions rise. Most of the data in that region can
be explained in terms of either small-$x$ limit of DGLAP equation or,
alternatively and more interestingly, by the BFKL equation\cite{bfkl},
which is also a linear equation but could be responsible for evolving
the system towards a higher gluonic density regime. However, the DIS
cross sections can not rise forever as powers of center of mass energy
$s$. This would violate unitarity\cite{froi} and pose other problems
discussed in Sect. 1. That means that at some very small $x$ the
hadronic structure functions have to undergo a significant qualitative
change of behavior, becoming a much slower varying functions of
$x$. The slow down of the growth of hadronic structure functions is
usually associated with non-linear effects in the quark and gluon
dynamics, such as parton recombination, which eventually balances the
parton splitting process as we have seen in the previous Section. The
partons in the hadronic wave function reach the state of {\it
saturation}.  The region of saturation of the structure functions is
represented in yellow in Fig. \ref{psat}. The scale $Q^2$
corresponding to the transition to the saturation region is different
for different values of Bjorken $x$, increasing with decreasing $x$
(increasing energy $s$). This scale is the {\it saturation scale}
$Q_s^2 (Y = \ln 1/x)$ given by \eq{eq:yvk:satsc}. If $Q_s^2 (x) \gg
\Lambda_{QCD}^2$ than the coupling constant inside the saturation
region would still be small allowing us to perform analytical
calculations inside of the saturation region\cite{mv} by employing
\eq{eqN}. 

It is possible that the recent experimental data obtained at HERA
\cite{epj:c7:609} contains evidence of saturation transition
in DIS on a proton at $x \approx 10^{-3} - 10^{-4}$ and $Q^2 \approx
2-4 \, GeV^2$, providing an experimental confirmation of the
theoretical ideas presented above.

\section{Open Questions}

Several question are still open in the field of saturation
physics. Here are some of the questions related to total cross
sections:

{\bf 1.} The large $N_c$ limit was instrumental in deriving
\eq{eqN}\cite{me}. Beyond it the systems of evolution equations in 
the existing approaches do not seem to
close.\cite{np:b463:99,jklw,ilm} Thus the question remains: can one
describe the high energy asymptotics of QCD beyond the large-$N_c$
limit?

{\bf 2.} NLO BFKL corrections\cite{nlo,nlo1} seem to be very large and
negative, posing several open questions for the linear evolution
equation.\cite{nlo2,ccs} Would these problems be cured if one
calculates the NLO correction to the full nonlinear evolution? 
Nonlinear effects become important before the NLO effects and may
modify the NLO corrections. Work on this project is under
way.\cite{bb}

{\bf 3.} Finally in this talk we have demonstrated how unitarization
happens in DIS cross sections via summation of fan diagrams of
\fig{fig:yvk:fan}. This mechanism would not work for onium-onium 
scattering cross section, even though a large $Q_s$ should exist in
the problem. One has to resum the pomeron loop
diagrams\cite{dip,np:b507:367} in order to understand the
unitarization of onium--onium cross section (and, correspondingly, the
proton-proton cross section). Resummation of pomeron loops is still an
important open question in high energy physics.

\section*{Acknowledgements}
I thank the organizers of {\it The Phenomenology of Large $N_c$ QCD}
workshop for such an interesting and timely meeting. This work has
been supported in part by the U.S. Department of Energy under Grant
No. DE-FG03-97ER41014 and by the BSF grant $\#$ 9800276 with Israeli
Science Foundation, founded by the Israeli Academy of Science and
Humanities.

\end{document}